# Simulation of beam-induced plasma in gas-filled rf cavities

Kwangmin Yu

*Computational Science Initiative, Brookhaven National Laboratory, Upton,
New York 11973, USA*

Roman Samulyak[*]

*Department of Applied Mathematics and Statistics, Stony Brook University, Stony Brook, New York 11794, USA
and Computational Science Initiative, Brookhaven National Laboratory, Upton, New York 11973, USA*

Katsuya Yonehara

*Fermi National Accelerator Laboratory, Batavia, Illinois 60510, USA*

Ben Freemire

*Northern Illinois University, DeKalb, Illinois 60115, USA*
(Received 22 June 2016; published 7 March 2017)

Processes occurring in a radio-frequency (rf) cavity, filled with high pressure gas and interacting with proton beams, have been studied via advanced numerical simulations. Simulations support the experimental program on the hydrogen gas-filled rf cavity in the Mucool Test Area (MTA) at Fermilab, and broader research on the design of muon cooling devices. SPACE, a 3D electromagnetic particle-in-cell (EM-PIC) code with atomic physics support, was used in simulation studies. Plasma dynamics in the rf cavity, including the process of neutral gas ionization by proton beams, plasma loading of the rf cavity, and atomic processes in plasma such as electron-ion and ion-ion recombination and electron attachment to dopant molecules, have been studied. Through comparison with experiments in the MTA, simulations quantified several uncertain values of plasma properties such as effective recombination rates and the attachment time of electrons to dopant molecules. Simulations have achieved very good agreement with experiments on plasma loading and related processes. The experimentally validated code SPACE is capable of predictive simulations of muon cooling devices.

## I. INTRODUCTION

Using muons is an attractive choice for realizing a multi-TeV lepton collider and producing a well-defined intense neutrino beam for neutrino experiments. Because muons are tertiary particles in the production process, the phase space volume of a muon beam needs to be shrunken to fit into the accelerator optics. Ionization cooling is a viable method to quickly cool down the beam temperature within the muon lifetime [1]. Muons propagate through an ionization material with strong magnetic focusing and lose their kinetic energy via ionization processes. The lost energy is immediately and adiabatically recovered by rf cavities. Better cooling efficiency is obtained with higher rf gradient. However, the achievable rf gradient is limited by the presence of a static magnetic field in a vacuum cavity because the density of dark current, which is a seed of electric breakdown in a rf cavity, is increased by magnetic focusing [2]. To resolve this problem, a high-pressure hydrogen gas-filled rf (HPRF) cavity was proposed [3,4]. Hydrogen gas serves the dual role: it buffers the dark current [5] and serves as the ionization material for the cooling process. A novel ionization cooling channel using the dual function cavity was proposed and its high cooling efficiency was demonstrated via simulations [6,7].

Experimental efforts have been made to characterize the HPRF cavity by using an intense proton beam in the Mucool Test Area (MTA) at Fermilab [8–10]. In particular, the plasma loading effect has been investigated [10]. Plasma loading takes place when a beam-induced plasma in the HPRF cavity absorbs the stored energy of the electromagnetic field in the cavity and causes reduction of the electric field. The plasma loading depends on the amount and mobility of charged particles in the gas. To decrease the plasma loading effect, adding a small amount of electronegative dopant to the hydrogen gas was proposed and successfully tested experimentally.

To better utilize experimental results and enable the design of a practical ionization cooling channel, a new

---
[*]roman.samulyak@stonybrook.edu

032002-1

algorithm to resolve plasma chemistry and plasma loading has been developed and implemented in the code SPACE [11]. The code reproduces complex plasma chemistry processes, including hydrogen recombination, electron attachment, and various recombination mechanisms that depend on plasma temperature.

A big challenge in such a plasma simulation is the need to simulate a very wide range of time scales, extending from the wake field characteristic time of picoseconds to millisecond intervals typical for plasma neutralization processes. This problem was resolved by utilizing new algorithms that perform simulations of the particle beam and plasma at different computational time steps. The main purpose of this paper is to validate new simulation algorithms by comparing them with experimental results. The second goal is quantification of uncertain values characterizing plasma processes. Even though the recombination rates and the attachment time were measured experimentally, the range of measurements was restricted to a narrow region in the plasma equilibrium state, and the measured values contained significant uncertainties due to experimental errors and the use of indirect measurement methods [9]. Comparison of simulation results with experimental data helps to calibrate parameters characterizing plasma properties and extend their values beyond the current experimental range. In particular, expressions for recombination rates have been established via comparison with experiments at certain values of pressure in the HPRF cavity. These expressions enable predictive simulations at the same gas pressure values but at different values of the electric field or different types of particle beams. In particular, simulation studies of a practical ionization cooling channel operating with intense muon beams will be reported in a forthcoming paper.

## II. MODELS AND NUMERICAL ALGORITHMS FOR ATOMIC PHYSICS PROCESSES IN PLASMA IN HPRF CAVITY

### A. Plasma formation

As an intense proton beam propagates in the HPRF cavity, plasma is created by ionization of hydrogen gas molecules due to collisions with beam particles. Some generated electrons have enough kinetic energy to cause secondary ionization. The process is described as

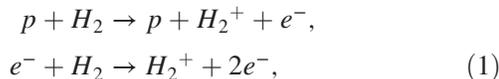

$$p + H_2 \rightarrow p + H_2^+ + e^-,$$
$$e^- + H_2 \rightarrow H_2^+ + 2e^-, \tag{1}$$

where $p$, $H_2$, and $e$ are a beam particle (proton), hydrogen molecule, and electron, respectively.

The number of electron-ion pairs $N_{pairs}$, produced in an elementary volume $dV$ of the cavity during time $dt$, is estimated based on the stopping power of the proton beam in hydrogen gas:

$$N_{pairs} = \frac{N_b}{W_i} \frac{dE}{dx} \rho h. \tag{2}$$

Here $(dE/dx)$ is a normalized stopping power of a beam particle per unit density of the absorber material, $\rho$, $h$, $W_i$, and $N_b$ denote the gas mass density, length of the volume $dV$ along the beam path, average ionization energy, and the number of beam particles that traverse $dV$ in time $dt$, respectively. In Eq. (2), $W_i$ accounts for both processes described in (1). The Bethe-Bloch formula [1] is used for calculating the stopping power.

At high pressures over 20 atm, typical for the HPRF cavity, the $H_2^+$ ions quickly form other clusters of hydrogen [12]:

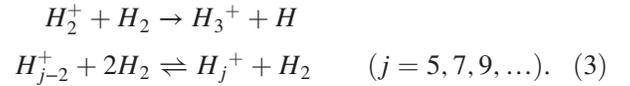

$$H_2^+ + H_2 \rightarrow H_3^+ + H$$
$$H_{j-2}^+ + 2H_2 \rightleftharpoons H_j^+ + H_2 \qquad (j = 5, 7, 9, \ldots). \tag{3}$$

### B. Recombination

In the rf cavity, filled initially with pure hydrogen gas, electrons created by ionization recombine with hydrogen clusters through either binary or ternary recombination processes:

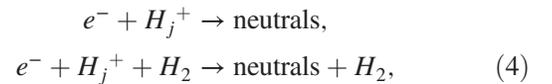

$$e^- + H_j^+ \rightarrow \text{neutrals},$$
$$e^- + H_j^+ + H_2 \rightarrow \text{neutrals} + H_2, \tag{4}$$

where $j = 3, 5, 7, \ldots$. At the HPRF conditions, ternary recombination is dominant [8]. The time evolution of the electron number density is given by the following equation

$$\frac{dn_e}{dt} = \dot{N}_e - \sum_j \beta_j n_e n_{H_j^+}, \tag{5}$$

where $j = 3, 5, 7, \ldots$, and $n_e$, $\dot{N}_e$, and $\beta_j$ denote the number density of electrons, the production rate of electrons, and the recombination rate of electrons and $H_j^+$, respectively. As individual recombination rates are unknown, an effective recombination rate $\beta_e$ for an averaged hydrogen ion cluster is used in our models and simulations. The effective recombination rate was measured in the MTA experiments at the equilibrium state of plasma (i.e. $dn_e/dt = 0$). In our model, we use the following fit for the effective recombination rate, applicable to transient processes in the plasma,

$$\beta_e = aX^b, \tag{6}$$

where $X = E/P$ is the ratio of electric field magnitude to the gas pressure. As described in Sec. III, values for the parameters $a$ and $b$ are obtained via comparison of HPRF simulations and various experimentally measures quantities, in particular, the plasma loading.

### C. Attachment and ion-ion recombination

When an electronegative gas such as oxygen is added to the hydrogen gas, a three-body attachment process takes place in the plasma, which is significantly faster than the electron-ion recombination [9]. The negative ions produced by the attachment process recombine with hydrogen ions. The governing equations are

$$\frac{dn_e}{dt} = \dot{N}_e - \beta_e n_e n_{H^+} - \frac{n_e}{\tau}$$

$$\frac{dn_{H^+}}{dt} = \dot{N}_e - \beta_e n_e n_{H^+} - \eta n_{H^+} n_{O_2^-}$$

$$\frac{dn_{O_2^-}}{dt} = \frac{n_e}{\tau} - \eta n_{H^+} n_{O_2^-} \tag{7}$$

where $\tau$, $\eta$, and $n_{O_2^-}$ are the attachment time, effective ion-ion recombination rate, and the number density of dopant gas ions, respectively. The averaged hydrogen ion cluster that represents the sum $\sum_j \beta_j n_{H_j^+}$ is denoted as $\beta_e n_{H^+}$.

As the recombination and attachment rates depend on the external field, they are functions of spatial coordinates and time. The attachment time and the ion-ion recombination rate have been measured experimentally, but only over a narrow range of rf field amplitudes. Based on measured values, simulations establish accurate fit functions for the attachment time and the ion-ion recombination rate.

### D. Numerical algorithms and their implementation

#### 1. Code SPACE

SPACE is a parallel particle-in-cell (PIC) code for the simulation of electromagnetic fields, relativistic particle beams, and plasmas based on the finite difference time domain (FDTD) method [13]. While the PIC method has been used for plasma simulations for a long time, the majority of PIC codes operate with either a constant number of plasma particles or simple mechanisms for plasma generation. The main feature of SPACE, relevant to HPRF simulations, is its ability to support complex atomic physics transformations, such as those described in the previous section. Some other notable examples of multiscale and multiphysics PIC simulations of plasmas can be found in [14,15].

In addition to the rigorous charge conservation algorithm [16], SPACE includes plasma chemistry, an efficient method for highly relativistic beams in nonrelativistic plasma, support for simulations in relativistic moving frames, and a special data transfer algorithm for transformations from moving to laboratory frames that resolves the problem of individual times of particles. The code also contains a highly adaptive and artifact-free particle method, called AP-Cloud, for solving the Vlasov-Poisson problems [17]. SPACE is written in C++ using a flexible and extendable modular structure [11]. A key component of the atomic physics module of the SPACE code is the variable representing number method for macroparticles that makes it possible to simulate ionization and recombination processes without causing artificial charge fluctuations. Here the representing number of a macroparticle means the number of real particles that are numerically approximated by this macroparticle. The algorithm also optimally resolves the change of plasma density by orders of magnitude. Details of the plasma chemistry module of SPACE are presented in Appendix, Secs. A 1 and A 2.

Another computational difficulty is the presence of different time scales, as in the case of plasmas interacting with relativistic particle beams. Evolution of atomic physics processes may be orders of magnitude longer compared to passing times of short relativistic bunches. The use of different time step sizes for the beam and plasma dramatically increased the code performance and enabled simulations of long physical times.

#### 2. Modeling and implementation of plasma loading

Due to high frequency collisions with neutral gas molecules, plasma electrons and ions reach their equilibrium drift speeds in a short period of time, and under the action of external rf fields [8]. The average energy loss by one plasma particle pair during one rf cycle, called $dw$, was introduced in [8,18]. Consider a collection of electrons and positive and negative ions in a small neighborhood of a point $(x, y, z)$ at time $t$. When the rf field is given by $E_0(x, y, z) \sin(\omega t)$, where $E_0(x, y, z)$ and $\omega$ denote the local amplitude and the angular frequency of the rf field, respectively, $dw$ for an effective electron-ion pair $j$ is

$$dw_j(t)$$
$$= q \int_0^t [p_e v_e + v_+ + (1 - p_e)v_-] E_0(x, y, z) \sin(\omega \tau) d\tau. \tag{8}$$

The effective electron-ion pair contains one positive ion, and a fraction of an electron $p_e$ and the corresponding fraction of the negative ion $1 - p_e$. $p_e = 1$ in pure hydrogen gas. Here subscripts "$e$", "$+$", and "$-$" denote electron, positive ion (hydrogen), and negative ion (dopant) respectively, and $q$, $T$, and $v$, are the particle charge, the rf field period, and the drift velocity of a charged particle. The dependence on the local coordinates $(x, y, z)$ is omitted in formulas below. When the drift velocity of charges has the form $v = \mu E_0 \sin(\omega t)$, where $\mu$ is the mobility of the charge, Eq. (8) becomes

$$dw_j(t)$$
$$= q \int_0^t [p_e \mu_e + \mu_+ + (1 - p_e)\mu_-] E_0^2 \sin^2(\omega \tau) d\tau. \tag{9}$$

In numerical simulations, $dw$ is computed in nodes of the PIC mesh, as part of the variable representing number algorithm described in the previous section. The mobility or drift velocity of electrons in pure hydrogen gas has been well studied. It is represented as a function of $X = (E/P)$, the ratio of electric field magnitude to the gas pressure [19,20]. Numerical simulations of SPACE use an interpolation function based on the drift velocity measurement in [20]. The dominant ion clusters in the plasma are $H_5^+$ or larger. Since experimental data on the effective mobility of hydrogen ion clusters contain some uncertainty, calibration simulations tested various mobility values. On the other hand, measured values for the mobility of dopant oxygen ions, formed by the electron attachment, are relatively accurate. Simulations used measurement values reported in [9,21]. Even if the velocity of ions is so small that the initial distribution of the ion electric field remains practically unchanged at later times, justifying the approximation of stationary ions, the ion mobility may still have some effect on the plasma loading of the cavity. In simulations, computed values for the number density of ions by ionization / attachment / recombination processes are combined with values of ion mobility to compute the ion component of $dw$.

Another factor contributing to the plasma loading simulations is the distribution of the rf field ($E_0(x, y, z)$) in the cavity, studied in [9]. The rf field amplitude significantly changes in the longitudinal direction, while the change in the radial direction is negligibly small within 2–3 mm, which is bigger than the plasma column. Therefore, the external rf field amplitude was approximated in the SPACE code as a quadratic polynomial in the longitudinal coordinate. Figure 1 shows normalized $E_z$ values in the HPRF cavity along the axis and at $r = 2.29$ mm [9], and the corresponding polynomial fit in the SPACE code. The difference between all lines is less than 2%.

The change of the external rf field due to plasma loading can be approximated by an LRC resonant circuit formula [10,22]. In particular, the rf power dissipation can be described as

$$P = \frac{V(t)[V_{\max} - V(t)]}{R} - CV(t)\frac{dV(t)}{dt}, \quad (10)$$

where $P$, $R$, $C$, $V_{\max}$, $V(t)$ are the rf field power, the shunt impedance, the cavity capacitance, the magnitude of the external rf voltage, and the instantaneous voltage value, respectively [8,22]. In the simulations, the total power of plasma loading is computed at each time step as a sum over all computational nodes $N$ of the PIC mesh

$$P(t) = \sum_{j=1}^{N} dw_j(t). \quad (11)$$

This value is used to compute the rf field voltage $V(t)$ by integrating the ordinary differential equation

$$\frac{dV(t)}{dt} = \frac{V_{\max} - V(t)}{RC} - \frac{P(t)}{CV(t)}, \quad (12)$$

and comparing it with experimentally measured values.

## III. SIMULATION RESULTS

### A. Hydrogen gas-filled rf cavity

In the rf cavity filled with pure hydrogen gas, ionization and electron-ion recombination are the main processes in the plasma. The main physics parameters are described in Table I.

The computational domain setting is shown in Fig. 2. Proton bunches of 2 mm radius and 25 cm length are injected into the computational domain with the time spacing of 5 ns. The vacuum gap between the computational domain boundary and the rf cavity wall is used to eliminate the effect of the numerical boundary condition.

Using the effective recombination rate measured at the MTA [23], simulations established accurate fit functions for

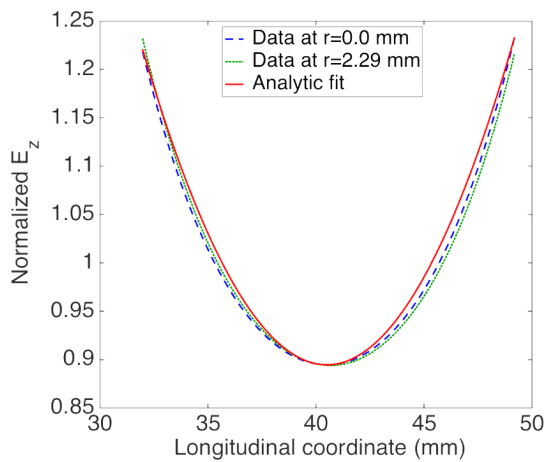

FIG. 1. Normalized $E_z$ distributions in HPRF cavity along the axis (blue dashed line) and at radius $r = 2.29$ mm (green dotted line), and its analytic fit in SPACE code (solid red line).

TABLE I. Parameters of HPRF cavity at 100 atm of pure hydrogen gas.

| Parameters | Values |
|---|---|
| Kinetic Energy of beam | 400 MeV |
| Beam Length | 25 cm |
| Beam Radius | 2 mm |
| $H_2$ gas pressure | 100 atm |
| dE/dx | 6.33 MeV cm$^2$/g |
| Average Ionization Energy | 36.2 eV |
| External Electric Field | 20.46 MV/m |
| (Frequency) | (801.6 MHz) |
| Population per Bunch | $2.23 \times 10^8$ |
| Bunch Spacing | 5 ns |
| Total Number of Beam Bunches | 2000 |

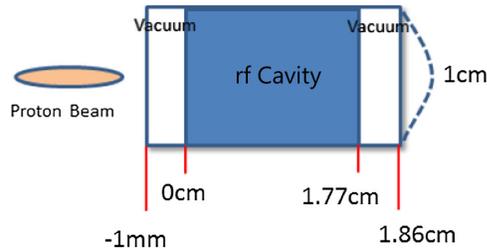

FIG. 2. Schematic diagram of the computational domain of the HPRF simulations.

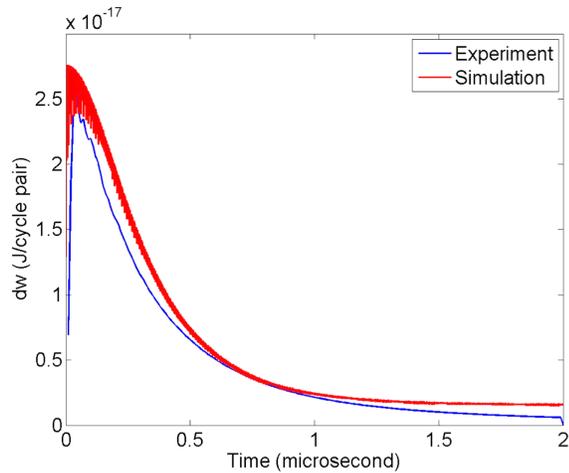

FIG. 3. Comparison of simulated and experimental values of $dw$ for HPRF cavity filled with pure hydrogen gas at 100 atm. Simulation $dw$ values represent average values over all macro-particles and averaged in time over one rf cycle.

the recombination rate and particle mobilities that result in excellent agreement with experimentally measured quantities. In particular, the fit obtained for the recombination rate is $\beta = 1.5 \times 10^{-10} \ X^{-1.2} \ (\text{cm}^3/\text{s})$ where $X = E/P$

(MV/m/psi). Figures 3, 4, and 5 show good agreement of simulations and experiment in terms between $dw$ and the plasma loading of the HPRF cavity. Figure 4 shows that the simulated magnitude of the electric field is within experimental errors denoted by error bars. The error bars include 10% measurement errors and standard deviations of 8 data points. After the beam is turned off at 10 $\mu$s, the external electric field recovers [Fig. 4(b)] as the number of electrons [Fig. 6(b)] decreases via recombination.

The equilibrium state of the plasma, when ionization is balanced by recombination, is achieved between 2 $\mu$s and 10 $\mu$s [see Figs. 4(b), 5(b), and 6(b)]. The electron-ion recombination rate was experimentally measured only in this region [9] while the fit function, established in simulations, is applicable to the transient region as well. Figure 4b shows that the magnitude of the external rf field is reduced by the factor of 7 at equilibrium. Such a large reduction is explained by a slow electron-ion recombination rate, resulting in high values of the plasma density.

The thick (red) band shown in Figs. 3, 5, and 6 is caused by high frequency oscillation of the corresponding quantities in simulations. These oscillations are caused by the structure of the beam, in particular the recombination processes in 5 ns intervals between bunches. In Fig. 5, the measurement of power is noisy, especially after the beam off time (10 $\mu$s) shown in Fig. 5(b).

### B. Hydrogen gas with dry air dopant

When an electronegative gas dopant is added to the hydrogen gas in the HPRF cavity, the electron attachment to dopant molecules and ion-ion recombination take place in addition to the electron-ion recombination process. The main parameters are described in Table II.

The attachment time and the ion-ion recombination were measured in the MTA experiment, albeit in a narrow range.

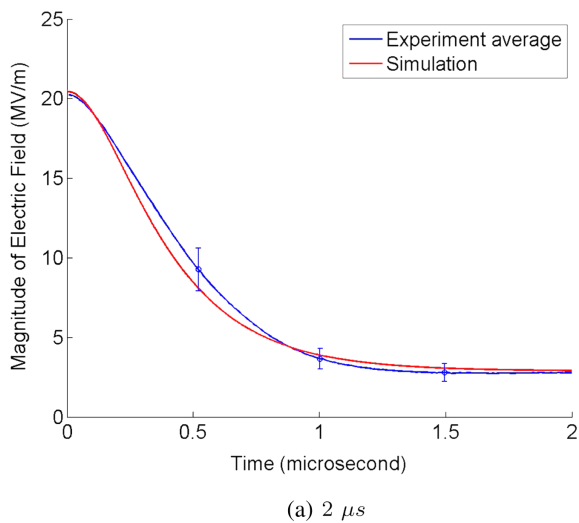

(a) 2 $\mu$s

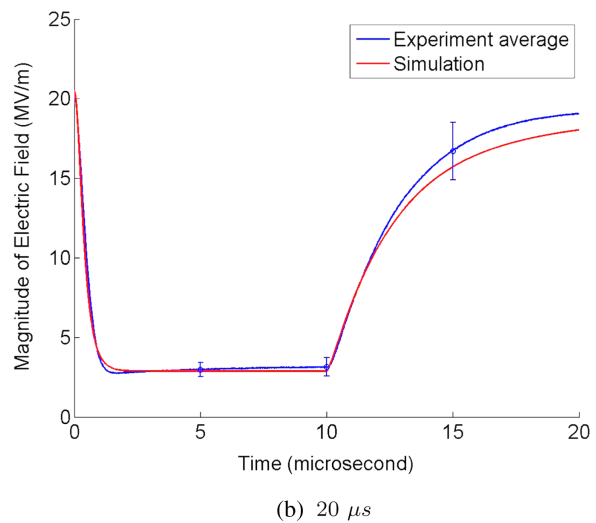

(b) 20 $\mu$s

FIG. 4. Comparison of simulated values of the magnitude of the external electric field with experimental values including error estimates in the HPRF cavity filled with pure hydrogen gas at 100 atm.

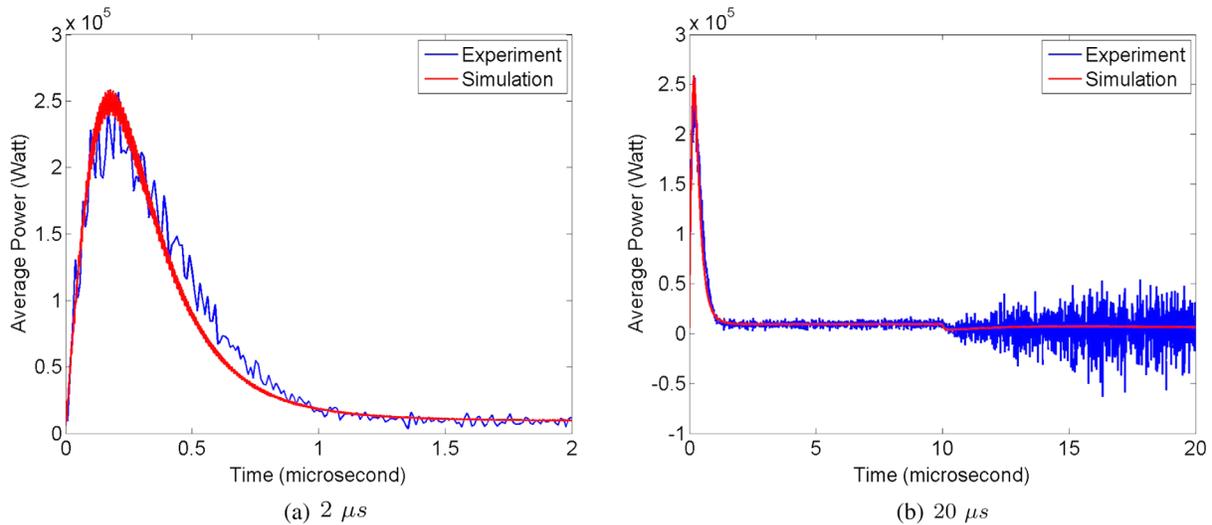

FIG. 5.   Comparison of values of power, averaged in time over one rf cycle, in the HPRF cavity filled with pure hydrogen gas at 100 atm obtained from experiments and simulations using formula (11).

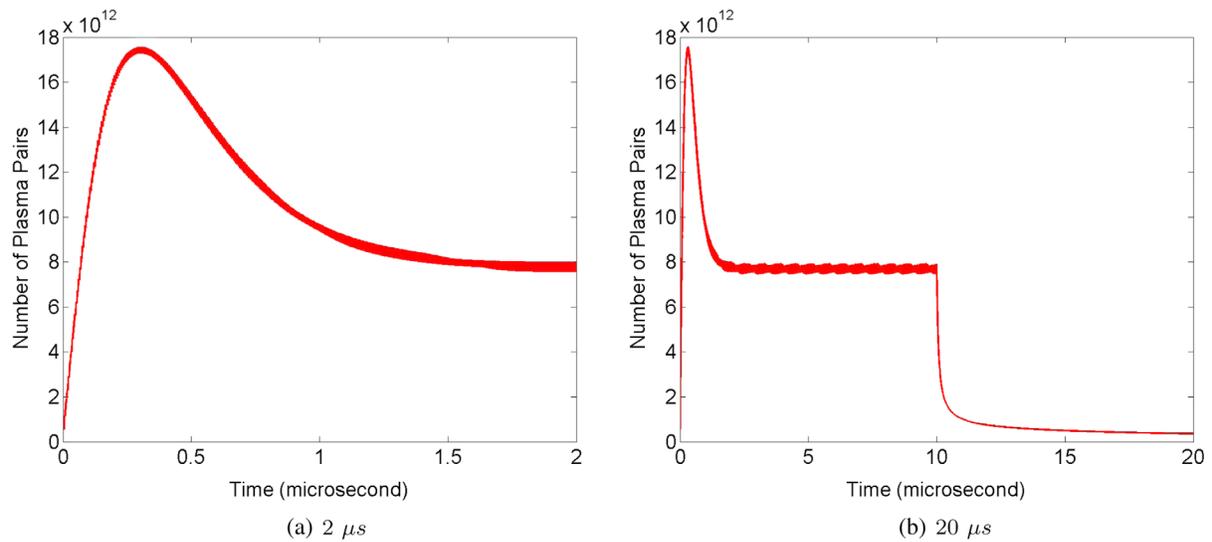

FIG. 6.   Simulated values of number of electrons in the HPRF cavity filled with pure hydrogen gas at 100 atm.

Similar to the pure hydrogen gas case, simulations achieved good agreement with experimentally measured quantities characterizing the plasma loading by finding the best fit functions. In particular, $\tau = 4.0 \times 10^{-7} \; X^{1.0}$ (s) and $\eta = 1.6 \times 10^{-10} X^{-1.0}$, where $X = E/P$ (MV/m/psi), were used.

TABLE II.   Parameters of HPRF cavity at 20.4 atm of hydrogen gas with 1% dry air. Only quantities with different values compared to Table I are shown.

| Parameters | Values |
| --- | --- |
| $H_2$ gas pressure (Dopant) | 20.4 atm (1% dry air) |
| External Electric Field | 8.84 MV/m |
| (Frequency) | (808.4 MHz) |
| Population per Bunch | $1.61 \times 10^8$ |

Plasma electrons become attached to dopant molecules in a short period of time. Figures 7(b), 7(c), and 7(d) show that the densities of hydrogen ions and dopant ions exceed the electron density by a factor of 50 even at 1 $\mu$s, and this ratio increases with time. We would like to note that quantities shown in Fig. 7(b), 7(c), and 7(d) fluctuate with the beam period of 5 ns. To eliminate this effect, all quantities are shown at the same phase of the proton beam, in particular when the proton bunch is in the center of the cavity.

In general, the effect of plasma loading of the HPRF cavity is governed by three processes: the electron attachment to dopant, the electron-ion recombination, and the ion-ion recombination. During the injection of the proton beam, the first process is dominant at early times, defining the initial slope of the external rf field magnitude in

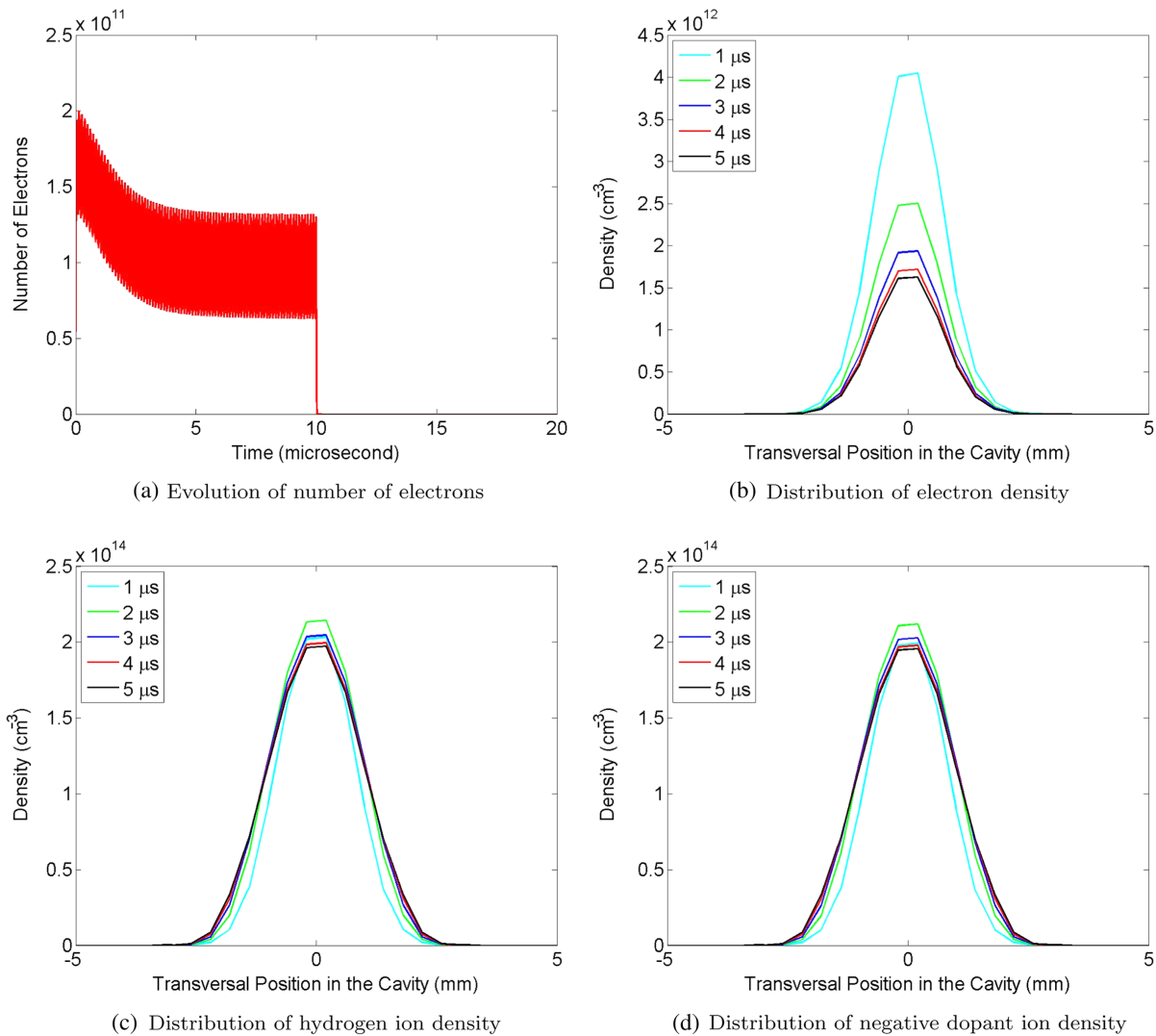

(a) Evolution of number of electrons

(b) Distribution of electron density

(c) Distribution of hydrogen ion density

(d) Distribution of negative dopant ion density

FIG. 7.   Charges in the HPRF cavity filled with 20.4 atm hydrogen gas and 1% dry air dopant. (a) Evolution of the total number of electrons in the cavity, (b), (c), and (d) show distribution of electrons, hydrogen ions, and dopant ions, respectively, in the center of the HPRF cavity.

Fig. 8(a). Both the attachment and ion-ion recombination are important after 0.5 $\mu$s, and the ion-ion recombination plays a much bigger role compared to the electron-ion recombination. After the beam is turned off [see Fig. 7(a)], the electron density rapidly decreases through attachment, and only the ion-ion recombination is effectively responsible for the recovery of the external rf field magnitude. Fluctuations of experimental values of the power in Fig. 8(b) are explained by measurement resolution of the rf signal. There is noise in the rf signal record, which propagates to the power in Fig. 8(b) impacting the $\frac{dV(t)}{dt}$ term in Eq. (10). But the simulation result of the power shows similar tendency with the experiment in Fig. 8(b). Figure 8(a) shows that the external electric field magnitude is reduced only by a factor of 1.7 at the equilibrium. This is a great improvement compared to the pure hydrogen case: simulations with pure hydrogen at the same conditions

(20.4 atm) show that the rf field magnitude at equilibrium is reduced by a factor of 44.

## IV. CONCLUSIONS

Models and algorithms for plasma dynamics in the HPRF cavity have been developed and implemented in the code SPACE. Numerical studies of the HPRF cavity have been performed, compared with experiments in the Fermilab MTA facility, and a very good agreement has been obtained. Dominant effects of the plasma dynamics in the HPRF cavity have been quantified in numerical simulations and previous analytical studies. They showed that ionization electrons are immediately thermalized by interacting with neutral particles and follow the conventional electron transport model. As reported in the experimental paper [8], electron mobility is significantly reduced by a pressure effect. Electron capture by an electronegative dopant is also

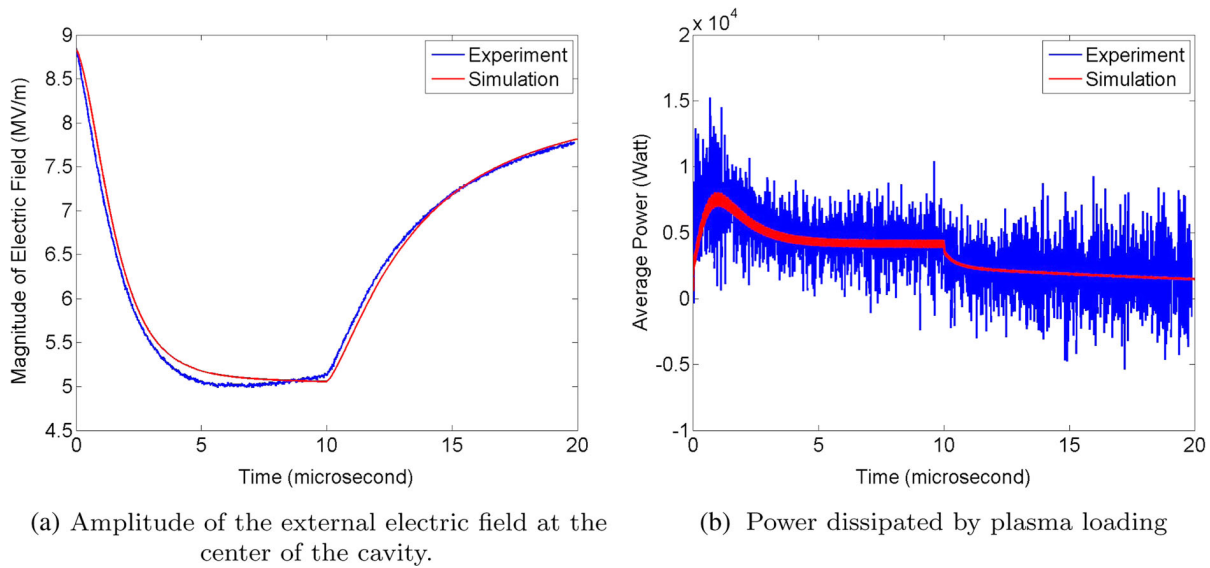

(a) Amplitude of the external electric field at the center of the cavity.

(b) Power dissipated by plasma loading

FIG. 8. Comparison of simulations and experiments for HPRF cavity filled with 20.4 atm hydrogen gas with 1% dry air dopant.

explained by the conventional three body model. It concludes that the electron capture time can be much shorter than a nanosecond. The spatial distribution of plasma is reproduced accurately by the code SPACE.

On the other hand, simulations show a very strong reduction of the external rf field magnitude in equilibrium for pure hydrogen plasma: the field is reduced by a factor of 7 at the pressure of 100 atm, and by the factor of 44 at 20.4 atm. The larger reduction of the electric field at low pressure is due to smaller recombination rates and, therefore, higher electron density. If a 1% dry air dopant is added to the hydrogen gas, the reduction of the rf field is greatly mitigated: the reduction factor at 20.4 atm is only 1.7. Simulations have achieved a very good agreement with experiments on plasma loading and related processes.

Simulations also contributed to a better understanding of plasma properties. In a series of simulations and their comparison with experimentally measured quantities characterizing the plasma loading process, several uncertain properties of the plasma, such as effective recombination rates and the attachment time of electrons to dopant molecules, have been quantified and accurate fit functions for these quantities, valid over a wide range of electric field values and the same pressure values, have been found. Validated using HPRF experimental data, the SPACE code is capable of predictive simulations at different electric field values and, more importantly, different particle beams (different ions or muons). In our current work, SPACE is used for the study of plasma loading in a practical ionization cooling channel operating with intense muon beams.

## ACKNOWLEDGMENTS

This research has been partially supported by the DOE Muon Accelerator Program. This manuscript has been authored in part by Brookhaven Science Associates, LLC, under Contract No. DE-SC0012704 with the U.S. Department of Energy.

## APPENDIX: IMPLEMENTATION OF ATOMIC PHYSICS ALGORITHMS IN SPACE

### 1. Generation of plasma macroparticles

Two algorithms for the dynamic generation of plasma have been implemented in SPACE. In this section, we describe the first algorithm that dynamically creates plasma macroparticle pairs and the second algorithm that changes the representing number of macroparticles is described in the next section. Consider an example of neutral gas ionization by a high energy particle beam. As each beam particle passes through the gas, it loses energy and ionizes the medium by creating electron-ion pairs. The amount of energy lost by the beam through ionization processes, or the beam stopping power, is described by the Bethe-Bloch equation [1]. This process is directly implemented in the code: the energy loss of every beam macroparticle is computed in real time, and a macro-electron-ion pair is numerically created when the beam particle energy loss exceeds the ionization energy. Each pair of electron and ion macroparticles is created in the same spatial location to satisfy the initial local charge neutrality. The mobility of ions is often very low throughout the simulation and the motion of ions can be ignored. As stationary particles have no effect on the solutions of the Maxwell equations, we need to create only electron macroparticles with zero initial electric field (due to plasma neutrality) in such a case. The initialization of only electrons with no electric field is equivalent to the presence of stationary positive ions by the conservative property of the numerical algorithm. As a result, the electric field due to stationary ions is present in simulations. The schematic of processes is shown in Fig. 9.

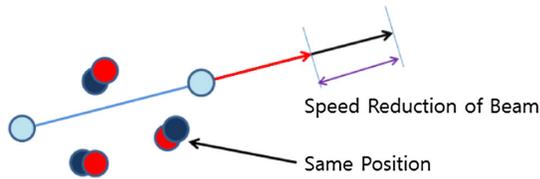

FIG. 9.   Schematic diagram of the ionization algorithm.

As the beam passes through a dense gas, beam particles also undergo scattering processes. The scattering of beam particles affects the ionization cooling and must be accounted for during the reacceleration phase of the beam. However, the scattering process do not change rates of atomic processes. As the study of such processes and their role in the performance of rf cavities is the primary interest of the paper, the scattering is neglected in the numerical model.

### 2. Variable representing number of macroparticles

In many applications, the plasma density can change by several orders of magnitude during physically relevant times of interest. The method described in the previous section may lead to numerical difficulties due to the extreme increase of the number of macroparticles in high density plasma regions while still achieving poor accuracy in low density regions represented by a small number of macroparticles. This method also leads to numerical oscillations caused by recombination processes. Initially overlapping electron and ion macroparticles, created by the ionization process, become spatially separated by dynamic processes at a later time. When a recombination event occurs at some point in space and time, due to the computed probability of recombination, the two closest but spatially separated charges must be eliminated, causing numerical oscillations.

This problem is effectively eliminated by using a variable representing number of macroparticles. In this algorithm, a preset cloud of massless, neutral plasma macroelectrons (with zero representing number) is created at the initial simulation time with a number density sufficient for numerical accuracy. At a later time, such macroparticles are charged proportionally to ionization processes by increasing their representing number, and decreasing the representing number proportional to the recombination rate. The ionization and recombination rates are computed on the PIC mesh, and the corresponding changes are interpolated to the location of all macro-electrons and are kept on PIC mesh nodes for ion components (as ion particles are not physically present). This algorithm keeps the number of computational macroparticles at the optimal level, removes numerical oscillations caused by recombination processes, and eliminates a need for a complex "bookkeeping" algorithm that records the lifetime of every plasma macroparticle and calculates the probability of its recombination or attachment to a dopant molecule.

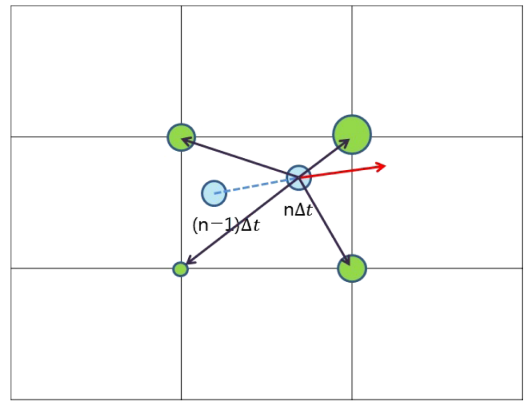

FIG. 10.   Schematic diagram illustrating ionization algorithm with variable representing number of plasma macroparticles.

Figure 10 shows the schematic description of the stopping power computation by a particle beam. By the movement of a beam particle (blue), its energy loss in gas is estimated and distributed to the FDTD mesh (green). At the same time, the energy loss of the beam particle is counted and used to update the velocity of the beam particle. After that, the number of new plasma ionization and recombination events is computed on the mesh. Their difference in each mesh point is interpolated to the plasma macroparticles contained within the domain of dependence of the interpolation scheme and used for changing of their representing number.

---